\documentclass[submission,Phys]{SciPost}
\pdfoutput=1
\usepackage{amsmath,amssymb,mathtools,xspace}
\usepackage{booktabs,multirow,tabularx,slashed}
\usepackage{hyperref}
\usepackage{color,xcolor}
\usepackage[normalem]{ulem}
\usepackage{enumitem}
\usepackage{braket,stackrel}
\usepackage{tikz}
\usetikzlibrary{shapes.geometric,arrows}
\makeatletter
\@ifundefined{pdfoutput}{}{\DeclareGraphicsRule{*}{mps}{*}{}}
\makeatother

\makeatletter
\DeclareRobustCommand*{\bfseries}{%
   \not@math@alphabet\bfseries\mathbf
   \fontseries\bfdefault\selectfont
   \boldmath
}
\makeatother

\hypersetup{colorlinks=true,
        linkcolor={blue!30!blue},
        citecolor={blue!50!blue},
        urlcolor={blue!30!blue}
         }

\definecolor{Gcolor}{HTML}{3b528b}
\definecolor{Dcolor}{HTML}{e41a1c}

\tikzstyle{generator} = [rectangle, rounded corners, minimum width=3cm, minimum height=1cm,text centered, draw=Gcolor]
\tikzstyle{discriminator} = [rectangle, rounded corners, minimum width=3cm, minimum height=1cm,text centered, draw=Dcolor]
\tikzstyle{io} = [circle, trapezium left angle=70, trapezium right angle=110, minimum width=1cm, minimum height=1cm, text centered, draw=black]

\tikzstyle{process} = [rectangle, minimum width=1cm, minimum height=1cm, text centered, draw=black]
\tikzstyle{decision} = [rectangle, minimum width=1cm, minimum height=1cm, text centered, draw=black]

\tikzstyle{arrow} = [thick,->,>=stealth]
\usepackage{xcolor}

\def\lrgen{\text{LR}_\text{gen}}

\newcommand\one{\leavevmode\hbox{\small1\normalsize\kern-.33em1}}

\newcommand{\qqquad}{\qquad \qquad}

\newcommand{\gev}{\text{GeV}}

\def\slashchar#1{\setbox0=\hbox{$#1$}           
   \dimen0=\wd0                                 
   \setbox1=\hbox{/} \dimen1=\wd1               
   \ifdim\dimen0>\dimen1                        
      \rlap{\hbox to \dimen0{\hfil/\hfil}}      
      #1                                        
   \else                                        
      \rlap{\hbox to \dimen1{\hfil$#1$\hfil}}   
      /                                         
   \fi}

\begin{document}

\begin{center}{\Large \textbf{
How to GAN Higher Jet Resolution
}}\end{center}

\begin{center}
Pierre Baldi\textsuperscript{1},
Lukas Blecher\textsuperscript{2},
Anja Butter\textsuperscript{2},
Julian Collado\textsuperscript{1},
Jessica N. Howard\textsuperscript{3}, \\
Fabian Keilbach\textsuperscript{2},
Tilman Plehn\textsuperscript{2},
Gregor Kasieczka\textsuperscript{4}, and
Daniel Whiteson\textsuperscript{3}
\end{center}

\begin{center}
{\bf 1} Department of Computer Science, University of California, Irvine, US \\
{\bf 2} Institut f\"ur Theoretische Physik, Universit\"at Heidelberg, Germany\\
{\bf 3} Department of Physics and Astronomy, University of California, Irvine, US \\
{\bf 4} Institut f\"ur Experimentalphysik, Universit\"at Hamburg, Germany \\
keilbach@thphys.uni-heidelberg.de
\end{center}

\begin{center}
\today
\end{center}


\tikzstyle{int}=[thick,draw, minimum size=2em]

\section*{Abstract}
{\bf QCD-jets at the LHC are described by simple physics
  principles. We show how super-resolution generative networks can
  learn the underlying structures and use them to improve the
  resolution of jet images. We test this approach on massless QCD-jets
  and on fat top-jets and find that the network reproduces their main 
  features even without training on pure samples. In addition, we show 
  how a slim network architecture can be constructed once we have 
  control of the full network performance.}

\vspace{10pt}
\noindent\rule{\textwidth}{1pt}
\tableofcontents\thispagestyle{fancy}
\noindent\rule{\textwidth}{1pt}
\vspace{10pt}

\newpage
\section{Introduction}
\label{sec:intro}

Recent innovations in machine learning (ML) have provided boosts to
many areas of particle physics. Ideas developed by the machine
learning community to solve tasks unrelated to physics often have
potential for applications within analysis of data in particle
physics, even beyond improvements to analysis of high-dimensional data
and speed improvements of first-principle simulations.  One such
recent development is the ability to enhance the resolution of
images\cite{790414,4270236}, by learning context-dependent general
rules that can be applied to specific observations to generate
estimates of higher-resolution versions of the observed images.
Hadronic jets produced in collisions at the Large Hadron Collider
(LHC) are obvious candidates for testing many ML-methods, as they are
measured in large numbers, they come with a simple theoretical
description, their complexity is balanced by their local detector
patterns, and they are an integral part of almost every LHC
analysis. In this paper, we apply super-resolution methods to LHC jets
for the first time, generating images of jets at significantly higher
resolution than the original observations.

The idea of using ML methods for exploring jets has a rich
history. Early jet classification studies date to the early
1990s~\cite{Csabai:1990tg,Lonnblad:1990bi}, and work has recently
gained momentum through applications of deep learning tools to
low-level jet observables organized as calorimeter
images~\cite{Cogan:2014oua,deOliveira:2015xxd,Baldi:2016fql,Komiske:2016rsd,Li:2020bvf,Carrazza:2019cnt}. This
approach can also be applied to the theoretically and experimentally
well-defined task of top-quark
tagging~\cite{Kasieczka:2017nvn,Macaluso:2018tck}. An alternative
approach to organizing calorimeter deposits as pixelated images is to
prepare a list of the 4-momenta of subjet
constituents~\cite{Almeida:2015jua,Butter:2017cot,Pearkes:2017hku,Erdmann:2018shi},
including recurrent neural networks inspired by language
recognition~\cite{Louppe:2017ipp, Guest:JetFlavorClassification} or
point
clouds~\cite{Komiske:2018cqr,Qu:2019gqs,Chakraborty:2020yfc,Bernreuther:2020vhm,Shlomi_2020}.
These various approaches have been compared in
detail\cite{Kasieczka:2019dbj}, revealing that their expected
performance in tagging hadronically-decaying top quarks is relatively
independent of the motivation and the architecture of the network.
Open questions include attempts to gain theoretical understanding of
the network's learned
strategy~\cite{Bradshaw:2019ipy,Kasieczka:2020nyd,Dolan:2020qkr,Faucett:2020vbu},
the stability with respect to detector
effects~\cite{Kasieczka:2018lwf,Kasieczka:2020yyl}, treatment of the
uncertainty~\cite{Bollweg:2019skg,Kasieczka:2020vlh}, extension to a
wide range of inputs~\cite{Qu:2019gqs}, and anomaly
detection~\cite{Heimel:2018mkt,Farina:2018fyg,Bernreuther:2020vhm,Cheng:2020dal}.

The first of these open questions inspires us to search for ways to apply machine learning to improve experimental jet measurements, by combining the basic rules of jet physics with the specific information of an observed jet.  Independent of the nature of a given jet, its
physics is described by relatively few ingredients, most notably collinear and soft QCD
splittings, which can be measured at the LHC~\cite{Bieringer:2020tnw}. These
basic principles can allow a super-resolution
algorithm\cite{790414,4270236} to accurately estimate the higher-resolution information that led to the observed results.
 Super-resolution algorithms are widely used in image
applications~\cite{Yang2014SingleImageSA,wang2020deep}, including those which use convolutional neural networks
CNNs~\cite{wang2015deep}. They can be combined with generative
networks~\cite{goodfellow2014gan,Butter:2020tvl}, which can describe
jets~\cite{shower,locationGAN,monkshower,juniprshower,Dohi:2020eda, lu2020sarm} and LHC
events~\cite{dutch,gan_datasets,DijetGAN2,gan_phasespace} and have the
potential to increase the speed of LHC event generators significantly~\cite{Gao:2020zvv,Bothmann:2020ywa,Chen:2020nfb,Verheyen:2020bjw,Backes:2020vka}.
Such
super-resolution GANs~\cite{srgan,esrgan} have already been applied to
cosmological simulations~\cite{Kodi_Ramanah_2020,li2020aiassisted}.

A simple super-resolution task in jet physics is to improve the
resolution of a calorimeter image, using general QCD
patterns~\cite{DiBello:2020bas}. It raises the question of
\textsl{whether an up-sampled jet image can include more information
  than the original, low-resolution image}.  Naively, it seems that
the answer must be no, based on the same reasoning that motivates the
argument that a generative network cannot produce more information
than exists in its statistically limited training data set. However,
this argument fails to account for the implicit knowledge embedded in
the architecture of the network, which can contribute information in
the same manner as a functional fit~\cite{Butter:2020qhk}. A
super-resolution network applied to LHC jets combines the information
from the low-resolution image with QCD knowledge extracted from the
training data, for instance the underlying theoretical principles of
soft and collinear splittings combined with mass drop patterns. While
we will not attempt to quantify the added information (such an answer
will depend on individual applications), we will show that
super-resolution networks can enhance calorimeter images, and that
training on QCD-jets vs top-quark jets indicates that model
uncertainties for this application are small.
Our detailed study of super-resolution for jets follows similar studies in
Ref.~\cite{DiBello:2020bas} of single particles, on the way to wider applications of
super-resolution networks in particle physics. For example, such
networks can automatically test the consistency of a data set when
applied to different layers of a calorimeter. With an appropriate
conditioning, they can become elements of a tagging algorithm.
Up-sampling from calorimeter to tracker resolution can provide
consistency tests between charged and neutral aspects of an event and
can be turned into a new way of identifying and removing pile-up.
This is especially promising, as both sides of the up-sampling (low-resolution calorimeter data and high-resolution tracking data) are
present and thereby allow training from data only.

\section{Super-resolution GAN for Jets}
\label{sec:gan}

\paragraph{Jet images} The task for our super-resolution networks is
to generate a high-resolution (HR), super-resolved (SR) version of a
given low-resolution (LR) image. While it is ill-posed in a
deterministic sense, as many distinct HR images can correspond to a
single LR image, it is well-defined in a statistical sense.

To test and benchmark our new method we use the standard, public top-tagging 
data set~\cite{Butter:2017cot,Heimel:2018mkt}, which is also the basis of the community paper
on supervised top-tagging~\cite{Kasieczka:2019dbj}. It contains 
jet images from $t\bar t$ and QCD
di-jet events generated with \textsc{Pythia}~\cite{pythia} for a
center-of-mass energy of $\sqrt{s}=14$~TeV, with
\textsc{Delphes}~\cite{delphes} used to model the ATLAS detector
response, and with clustering and jet-finding done with
\textsc{FastJet}~\cite{fastjet1}. The fat anti-$k_\textrm{T}$
jets~\cite{anitkt} have a radius $R=0.8$ and
\begin{align}
  p_{\textrm{T},j}=550~...~650~\gev
  \qquad \text{and} \qquad
  |\eta_j|<2 \; ,
\end{align}
to have access to decent experimental resolution.  The jet images are
defined by pixel-wise $p_\textrm{T}$, with order of 50 active
pixels. This means that, for instance, images with $160\times160$
pixels have a sparsity of 99.8\%.  For the training of
super-resolution models, we provide paired LR/HR jet images, which are
generated by down-sampling the HR image. We use sum pooling on the jet
constituents as an approximation to reduced detector resolution before
we perform jet finding~\cite{cycleSR}.  After jet finding, we select
the hardest jet in each of the HR and LR images as a candidate pair,
rejecting the pair if either jet has fewer than 15 constituents.  To
ensure that the selected HR-clustered and LR-clustered jets correspond
to the same hard parton, we require the angular distance between the
two to be $\Delta R = \sqrt{ \Delta\eta^2 + \Delta\phi^2} < 0.1$. This
procedure defines paired HR and LR jet images, where the LR jet image
contains no information from the HR image.  We apply this procedure to
create LR-HR image pairs with down-scaling factors of 2, 4, and 8,
removing events that fail the requirement for any particular
resolution from all samples, which ensures that all jet samples
contain the same set of events.

\begin{figure}[t]
    \includegraphics[width=1.0\textwidth]{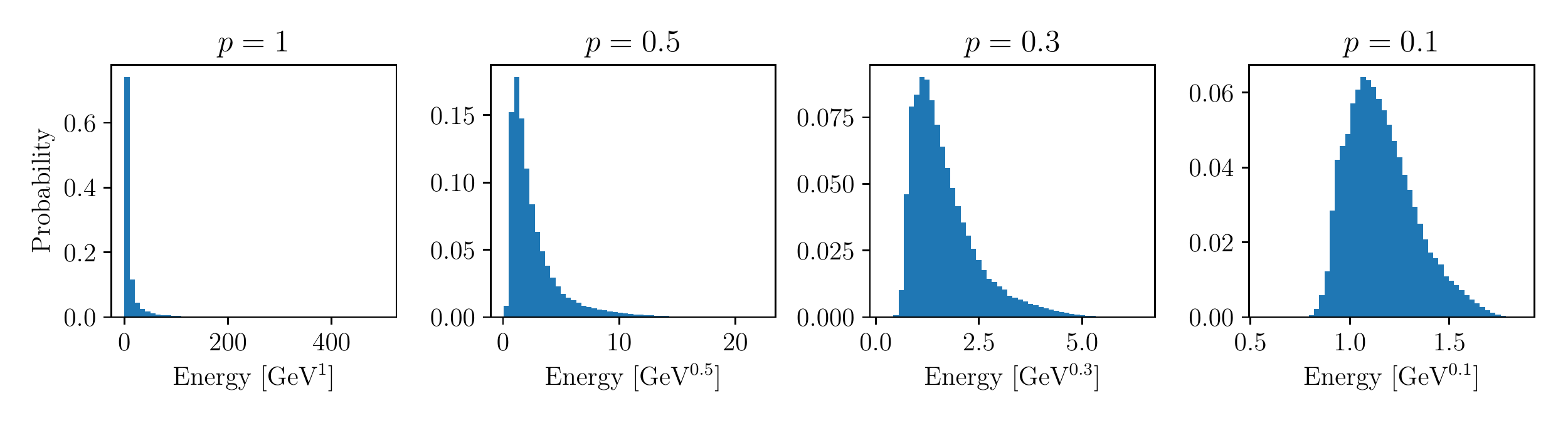}
    \caption{Distribution of energy deposition when pixel entries are raised by several
      different powers $E \to E^p$.}
    \label{fig:preprocess}
\end{figure}

There are multiple ways of normalizing jet images to be better suited
for machine learning. Such transformations do not retain the absolute
momentum, which may not be a problem for classification, but for our
purposes this information is needed. In Fig.~\ref{fig:preprocess}, we
show typical energy distributions after re-scaling the pixel entries
with a power $p$. Clearly, some kind of re-scaling is helpful to
enhance the otherwise extremely peaked spectrum. On the other hand, we
know that the low-energy radiation is largely noise, which means that
choosing $p$ too small is not helpful for the network to learn the
leading patterns. We find that $p=0.3$ is a good compromise, to be
combined with the original image $p=1$.

\begin{figure}[t]
    \centering
    \includegraphics[width=0.75\textwidth]{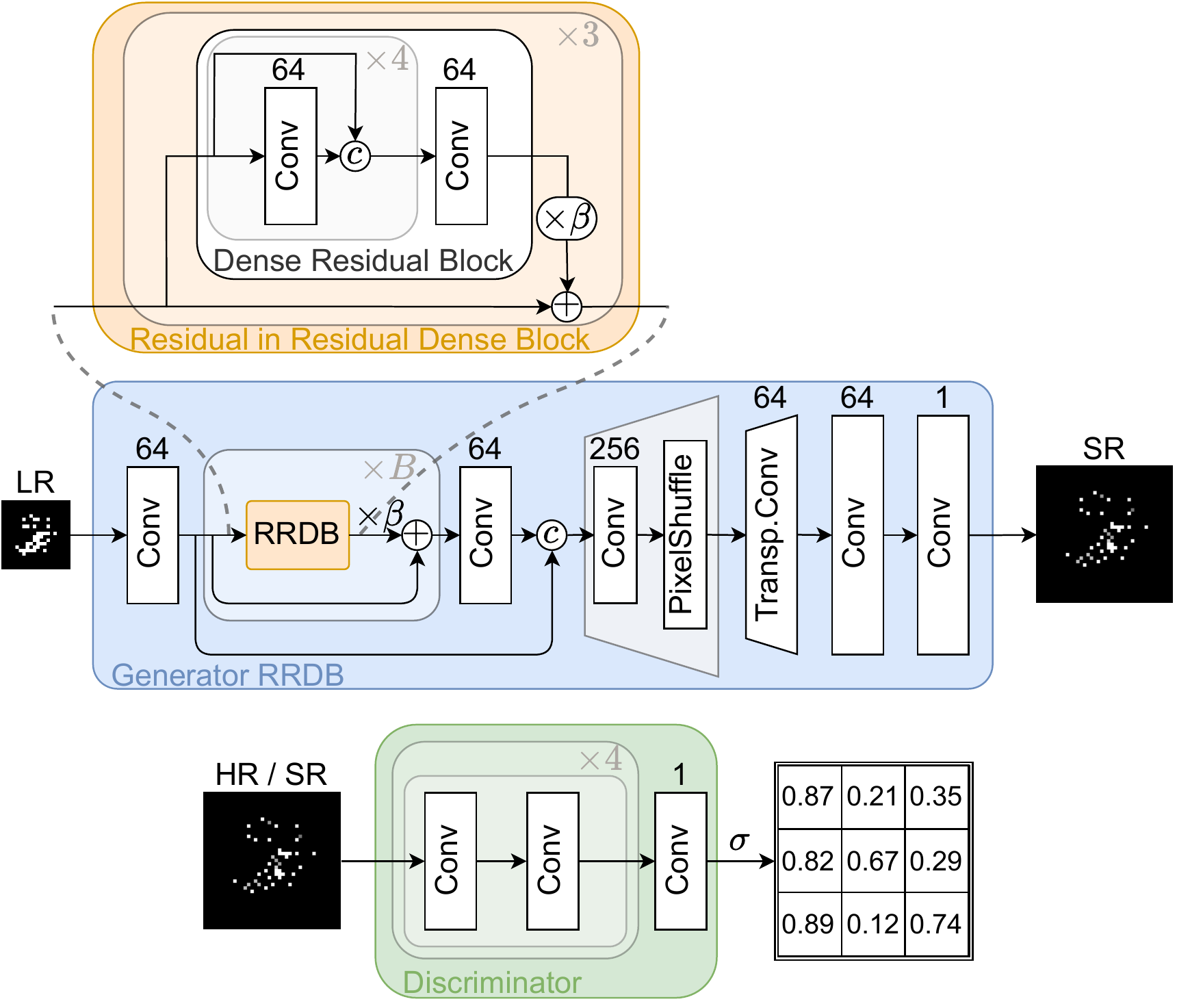}
    \caption{Architecture of modified generator network from ESRGAN (upper) and discriminator network modified
      from the SRGAN (lower).}
    \label{fig:network}
\end{figure}

\paragraph{Network architecture} In our jet image study, we use a
variant of the enhanced super-resolution GAN (ESRGAN)~\cite{esrgan},
illustrated in Fig.~\ref{fig:network}. To begin with, the generator
converts a LR image into a SR image using a deep residual fully
convolutional network. Its main element is the dense residual block
(DRB)~\cite{zhangResidualDense}, built out of consecutive
convolutional layers with ($3\times3$)-kernels, stride 1, padding 1, and
64 filters. The activation function is a LeakyReLU with $\alpha =
0.2$. The particularity of the DRB is that a layer receives the input
of all other layers in addition to the output of the previous layer.
This structure fuses all the feature maps inside the block. Three DRBs form a
residual-in-residual dense block (RRDB)~\cite{esrgan}, connected via
residual connections.

All convolutions in the generator preserve the spatial dimensions of
the input image. Following Fig.~\ref{fig:network}, the up-sampling
can be done by pixel-shuffle layers~\cite{pixelshuffle} or transposed
convolutions. Our generator up-samples by a factor of two in up to
three consecutive steps and works best if we alternate between
pixel-shuffle and transposed convolutions.  In the HR feature space,
there are two additional convolutional layers, one of which simply
scales the output by a fixed value.

The discriminator network is a relatively simple feed-forward
convolutional network with LeakyReLU activations, as proposed for the
SRGAN~\cite{srgan}. It uses blocks consisting of two convolutional
layers with a ($3\times3$)-kernel and padding 1. While the first
convolution of each block conserves the spatial dimensions, the second
layer halves it through a strided convolution. We link four of those
blocks and start with 64 filters, doubling the number of filters after
each block. We modify the original SRGAN structure by removing the
batch normalization layers and adding a gradient
penalty~\cite{gradientpenalty_improvedwgan,pix2pix,ragan}.  We cut off
the network before flattening, feeding it into a fully connected layer
and switching to a Markovian discriminator. Finally, we include a
second discriminator with exactly the same structure, such that the
full discriminator response is the sum of two discriminator networks.
For the second discriminator, we reset all weights after a fixed
number of batches.

\begin{table}[!b]
\centering
\begin{small}
\begin{tabular}{ccccccccccccc}
\toprule
& \#RRDB  & batch size & $ \beta $ & rescaling & $ \lambda_{\text{reg}} $ & $ \lambda_{\text{std}} $ & $ \lambda_{\text{pow}} $ & $ \lambda_{\text{HR}} $ & $ \lambda_{\text{LR}} $ & $ \lambda_{\text{adv}} $ & $ \lambda_{\text{patch}} $ & \text{reset interval} \\ \midrule
optimal & 10 & 15 &0.1&0.3 & 0.001 & 0.2           & 1             & 1        & 0.1       & 0.01    & 0.1 & 20k      \\
medium & 15 & 15 &0.1&0.3 & 0.001 & 1.2           & 1             & 1        & 0.1       & 0.05    & 0.1 & 20k      \\
\bottomrule
\end{tabular}
\end{small}
\caption{Sets of hyperparameters used for networks described in Fig.~\ref{fig:network}. Two sets are presented, one which optimized performance, and a second which which performed slightly worse. $\beta$ is the residual scale factor.}
\label{tab:hyper} 
\end{table}

\paragraph{Loss function} The SRGAN and ESRGANs
include a set of excess functionalities, such as perceptual loss which
can potentially improve the quality of the output. This loss combines
the adversarial loss from the discriminator with a content loss that
compares feature maps of a pre-trained image classification network.
The adversarial loss for a relativistic GAN trained on true events
($T$) to generate new events ($G$) is
\begin{align}
  L_\text{adv} =
  -\langle \log D \rangle_G
  &-\langle \log (1-D)\rangle_T \notag \\
  \text{with} \quad
  D_T =& \sigma \left(C_T - \langle C \rangle_G \right) \notag \\
  D_G =& \sigma\left(C_G - \langle C \rangle_T \right)
  \label{eq:adversarial}
\end{align}
where $\sigma$ is a sigmoid classifier function and $C$ is the
unactivated discriminator output. Compared to a standard 
adversarial loss, we have an additional term because $D_\textrm{T}$ depends on
the generated data $G$. The original content loss is not needed for
our purpose. Because our HR images should resemble the ground truth,
we add a $L_1$ loss between the SR and HR images. Our choice of $L_1$
over $L_2$ prevents blurring,
\begin{align}
  L_\text{HR}=L_1 \left(\text{SR},\,\text{HR}\right) \; .
  \label{eq:hrloss}
\end{align}
In return, because the LR image should correspond to the HR-jet, we
define a loss term that compares the model input with the down-sampled
model output pixel by pixel,
\begin{align}
  L_\text{LR}=L_1 \left(\sum_\text{pool}(\text{SR}),\,\text{LR}\right) \;.
  \label{eq:lrloss}
\end{align}
When we up-sample the LR-jet image by a factor $f$, we need to
distribute each LR pixel energy over $f \times f$ SR pixels. These $f
\times f$ pixels define a patch, and we encourage the network to
spread the LR pixel energy such that the number of active pixels
corresponds to the HR truth. This defines the loss term
\begin{align}
  L_\text{patch}=L_2 \left(\text{patch(SR)},\,\text{patch(HR)}\right)
  \label{eq:hitoloss}
\end{align}
This patch loss is strictly not necessary for the network, but with the
right choice of hyperparameters it balances the spread of constituents
in the jet image plane with the appearance of artefacts. 
The combined generator loss over the standard and re-weighted jet
images is then
\begin{align}
  L_G = \sum_{s\in\{\text{std},\,\text{pow}\}}
    \lambda_s \left(\lambda_\text{HR}\, L_\text{HR}
    + \lambda_\text{LR} \, L_\text{LR}
    + \lambda_\text{adv} \, L_\text{adv}
    + \lambda_\text{patch} \, L_\text{patch} \right) \; ,
    \label{eq:totalloss}
\end{align}
Our analysis included a careful scan over the hyperparameters $\lambda_j$. 
The metric we used to evaluate the performance of the models is the KL-divergence 
between the observables shown for instance in Fig.~\ref{fig:qcd-qcd}, 
most notably the energy of the hardest constituent and the average patches.

The GAN discriminator $D$ measures how close the generated data set
$G$ is to the true or training data $T$. In a relativistic average
GAN~\cite{ragan}, the discriminator is given by the probability of
a generated event being more realistic than the average true event, and vice
versa. It corresponds to the adversarial generator loss in
Eq.\eqref{eq:adversarial} but with switched labels,
\begin{align} =
  - \langle \log(1-D) \rangle_G
  - \langle \log D \rangle_T \; .
    \label{eq:discrganloss}
\end{align}
To this expression we add a gradient penalty for stabilization,
\begin{align}
    L_\text{reg} = \langle(\left\Vert \nabla_{X'}C(X')\right\Vert_2 - 1)^2 \rangle \; .
\end{align}
where $X'$ is a randomly weighted average between a real and generated
samples, $X' = \epsilon X_T + (1 - \epsilon) X_G$ and C(X') is the
unactivated discriminator output.

All hyperparameters are listed in Tab.~\ref{tab:hyper}. We use
\textsc{Adam}~\cite{Kingma:2014vow} for the optimization with
$\beta_1=0.5$~\cite{radford2015unsupervised} and $\beta_2=0.9$. The
learning rate is $\lambda=0.0001$. The training of a model typically
takes 50k-100k iterations.  

\begin{figure}[t]
    \centering
    \includegraphics[width=.5\textwidth]{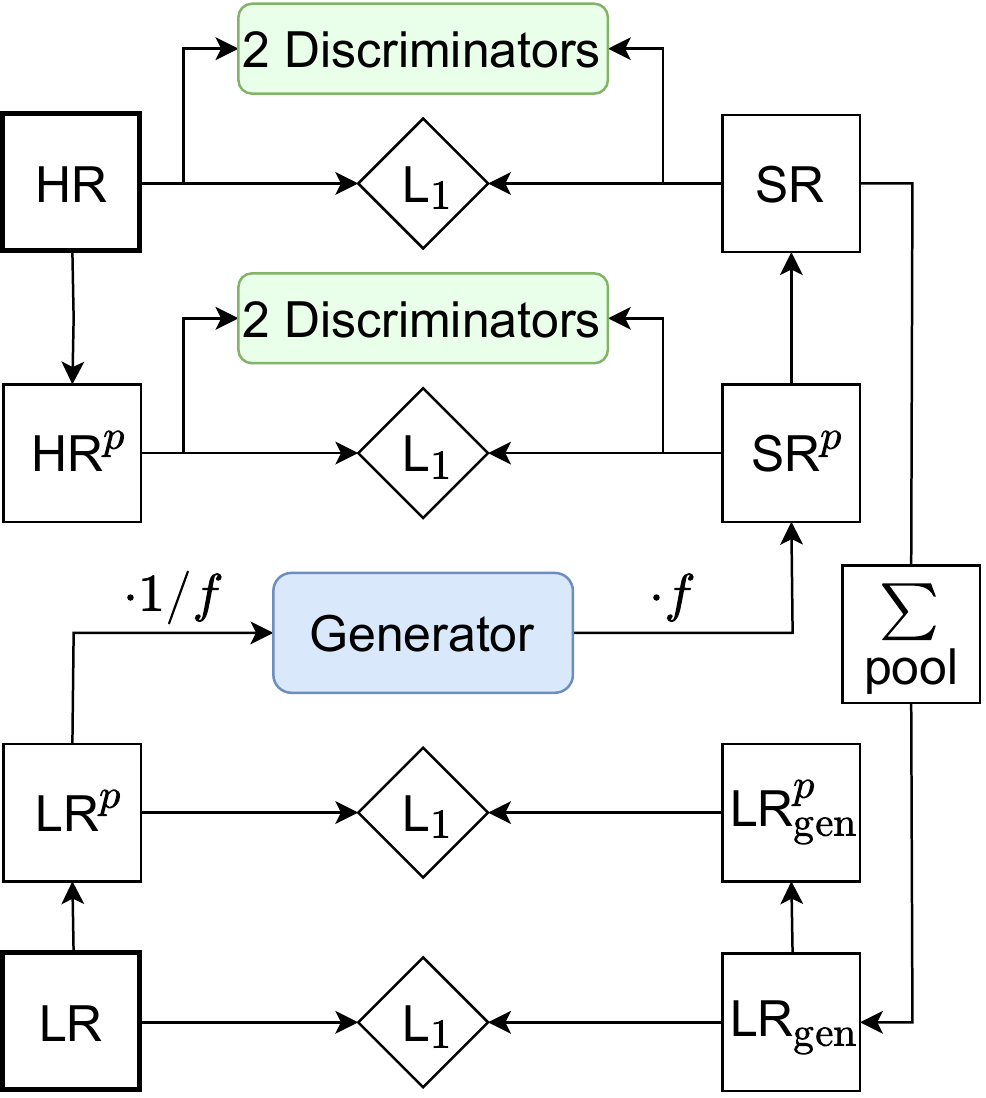}
    \caption{Training process for jet images. The generator and
      discriminator networks are shown in Fig.~\ref{fig:network}.}
    \label{fig:training}
\end{figure}

\paragraph{Training} The starting point of our training, illustrated in
Fig.~\ref{fig:training}, is the HR truth image, from which the LR
image is derived. All jet images are also raised to the power $p=0.3$
as a pixel-wise operation. We work with an up-scale factor
$f=2^3=8$. In that case, we  divide the LR image by the total
factor $f$ and feed it into the RRDB generator. Its output is divided
by the factor $1/f$ and gives the SR image raised to the power
$p$. This intermediate result is saved for the computation of
$L_\text{HR}$. For the SR output image we need to only take the
$p$\textsuperscript{th} root.  This SR image is sum-pooled back to its
LR version $\lrgen$ to compute the different generator loss terms.
Based on this set of LR, HR, and SR images, with and without a
$p$-scaling, we compute a set of $L_1$ loss contributions to the
generator loss, as well as the discriminator losses from the HR-SR
comparison.

\section{Up-sampling jets}
\label{sec:jets}

We benchmark the performance of the super-resolution algorithm for
both QCD jets and top-quark jets. QCD jets, which at the LHC arise
from massless partons, exist in large samples and are well described
by collinear and soft splittings. As an alternative, we use jets from
top-quark decays, which are significantly different, but can be
isolated experimentally from semi-leptonic top-quark pair production
and well-described theoretically via perturbative QCD.

We start with a set of HR-jet images with $160 \times 160$ pixels. We
down-sample each of these images to a corresponding LR image by a
linear factor $1/f = 1/8$ to an image of $20 \times 20$ pixels. For
the up-sampling, we apply three doubling steps using pixel shuffle,
transposed convolution, and another pixel shuffle. The pixel shuffle
has the advantage of encoding the full information from the feature
maps. It simply redistributes the information by transforming a large
number of channels, as usually arise after deep convolutions, into a
set of feature maps with fewer channels but larger spatial dimensions.
The transposed convolution takes into account local information
through a trainable kernel. After learning meaningful weights it can
help learning intricate, non-local patterns, which would be missed by
a global pixel shuffle. In the following, we first train and test a
network on QCD-jets, then on top-jets. To estimate the model
uncertainties, we apply networks trained on one class to the other
class.

To evaluate the quality of the information in our image-based results
in a physics context, we calculate an established set of jet
observables~\cite{Gallicchio:2010dq,Larkoski:2013eya,Thaler:2010tr,Kasieczka:2018lwf}
\begin{alignat}{9}
  m_\text{jet} &= \left( \sum_i p_i^\mu \right)^2
&  w_\text{pf} &= \frac{\sum_i p_{\textrm{T},i} \Delta R_{i, \text{jet}}}{\sum_i p_{\text{T},i}} \notag \\
  C_{0.2} &= \frac{\sum_{i,j} p_{\textrm{T},i} p_{\textrm{T},j} (\Delta R_{i,j})^{0.2}}{(\sum_i p_{\textrm{T},i})^2} \qqquad
& \tau_N &= \frac{\sum_k p_{\textrm{T},k} \text{min}(\Delta R_{1,k}, ... , \Delta R_{N,k})}{\sum_k p_{\textrm{T},k} R_{0}} \; .
\label{eq:jetobs}
\end{alignat}
The jet mass is the most relevant difference between pure QCD jets and
top decay jets.  The girth $w_\text{pf}$ essentially describes the
geometric extension of the hard pixels, while $C_{0.2}$ is the leading
pixel-to-pixel correlation. The subjettiness ratios $\tau_2/\tau_1$
and $\tau_3/\tau_2$ can distinguish between 2-prong and 3-prong decay
jets.

\subsection{Performance in QCD Jets}
\label{sec:jets_qcd}

\begin{figure}[t]
\centering 
\includegraphics[width=0.30\textwidth,page=2]{figures/qcd_M2}
\includegraphics[width=0.30\textwidth,page=3 ]{figures/qcd_M2}
\includegraphics[width=0.30\textwidth,page=4 ]{figures/qcd_M2}\\
\includegraphics[width=0.30\textwidth,page=5]{figures/qcd_M2}
\includegraphics[width=0.30\textwidth,page=6 ]{figures/qcd_M2}
\includegraphics[width=0.30\textwidth,page=7 ]{figures/qcd_M2}\\
\includegraphics[width=0.30\textwidth,page=8]{figures/qcd_M2}
\includegraphics[width=0.30\textwidth,page=9 ]{figures/qcd_M2}
\includegraphics[width=0.30\textwidth,page=10 ]{figures/qcd_M2}\\
\includegraphics[width=0.30\textwidth,page=1]{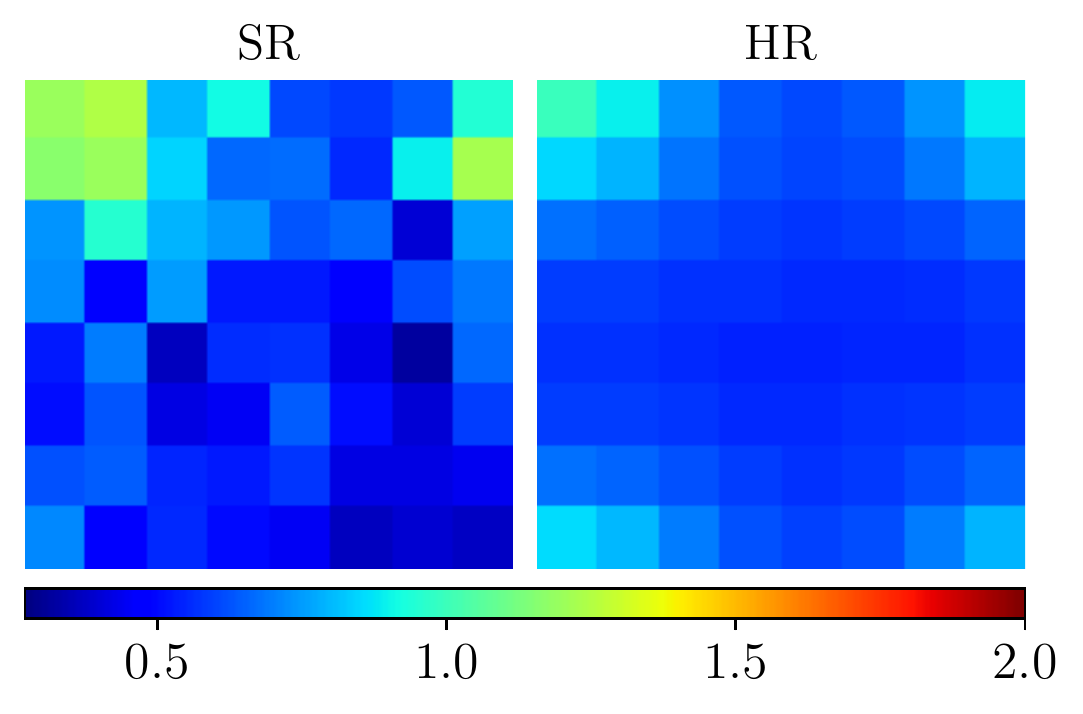}
\includegraphics[width=0.30\textwidth,page=11 ]{figures/qcd_M2}
\includegraphics[width=0.30\textwidth,page=14 ]{figures/qcd_M2}\\
\includegraphics[width=0.30\textwidth,page=15 ]{figures/qcd_M2}
\includegraphics[width=0.30\textwidth,page=12 ]{figures/qcd_M2}
\includegraphics[width=0.30\textwidth,page=13 ]{figures/qcd_M2} \\[-4mm]
\caption{Demonstration of the performance of a network trained on
  QCD-jets and applied to QCD-jets.  Top left are averages of the HR
  and SR images, followed by distributions of the square-root of the
  energy of leading pixels, sub-leading, etc. Also shown are average
  ($f \times f$)-patches for the SR and the HR images, and
  distributions of high-level jet observables, see text for
  definitions. The zero-bin in energy collects jets with too few
  entries.}
\label{fig:qcd-qcd}
\end{figure}

In an initial test, we train and test our super-resolution network on
the sample of QCD jets, which are characterized by a few central
pixels which carry most of the jet energy. In this case, it is
important to include down-sampled kinematic distributions in the
evaluation, to disentangle the central patterns.

In Fig.~\ref{fig:qcd-qcd} we compare the HR and SR images as well as
the true LR image with their generated $\lrgen$ counterpart. In
addition to average SR and LR images, we show the energy spectra for
the leading four pixels. This reveals how the LR image resolution
reaches its limits, because the leading pixel carries most of the
information. The sub-leading pixels are often harder for the HR image,
because the up-sampling often splits the hardest LR pixel. From the
7th leading pixel and beyond, we see an increasing number of empty
pixels, and above the 10th pixel the QCD jet largely features soft
noise. This transition is the weak spot of the SR network. While it
learns the underlying principles of QCD splittings for the hard pixels
and the noise patterns for the soft pixels, the mixed range around the
7th and 10th pixels indicates sizeable deviations. We also show the
average ($f \times f$)-patches for the SR and the HR images to confirm
that the spreading of the hard pixels works at the 20\% level.

\begin{figure}[t]
\centering
\includegraphics[width=0.30\textwidth,page=2]{figures/top_M2}
\includegraphics[width=0.30\textwidth,page=3 ]{figures/top_M2}
\includegraphics[width=0.30\textwidth,page=4 ]{figures/top_M2}\\
\includegraphics[width=0.30\textwidth,page=5]{figures/top_M2}
\includegraphics[width=0.30\textwidth,page=6 ]{figures/top_M2}
\includegraphics[width=0.30\textwidth,page=7 ]{figures/top_M2}\\
\includegraphics[width=0.30\textwidth,page=8]{figures/top_M2}
\includegraphics[width=0.30\textwidth,page=9 ]{figures/top_M2}
\includegraphics[width=0.30\textwidth,page=10 ]{figures/top_M2}\\
\includegraphics[width=0.30\textwidth,page=1]{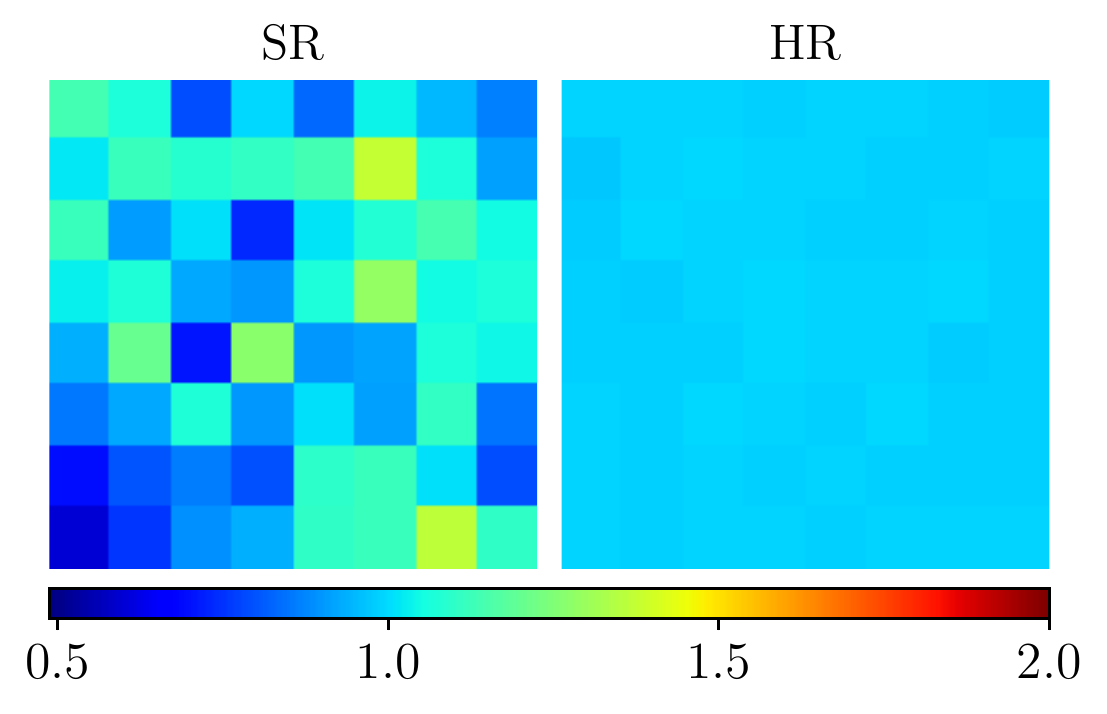}
\includegraphics[width=0.30\textwidth,page=11 ]{figures/top_M2}
\includegraphics[width=0.30\textwidth,page=14 ]{figures/top_M2}\\
\includegraphics[width=0.30\textwidth,page=15]{figures/top_M2}
\includegraphics[width=0.30\textwidth,page=12 ]{figures/top_M2}
\includegraphics[width=0.30\textwidth,page=13 ]{figures/top_M2} \\[-4mm]
\caption{Demonstration of the performance of a network trained on
  top-quark jets and applied to top-quark jets.  Top left are averages
  of the HR and SR images, followed by distributions of the
  square-root of the energy of leading pixels, sub-leading, etc. Also
  shown are average ($f \times f$)-patches for the SR and the HR
  images, and distributions of high-level jet observables, see text
  for definitions. The zero-bin in energy collects jets with too few
  entries.}
\label{fig:top-top}
\end{figure}

Again in Fig.~\ref{fig:qcd-qcd} we see that the jet mass peaks around
the expected 50~GeV, for the LR and for the HR-jet alike. Still, the
agreement between LR and $\lrgen$ on the one hand and between HR and
SR on the other is better than the agreement between the LR and HR
images. A similar picture emerges for the $p_\textrm{T}$-weighted
distance to the jet axis, the girth $w_\text{pf}$, which essentially
describes the extension of the hard pixels.  The pixel-to-pixel
correlation $C_{0.2}$ also shows little deviation between HR and SR on
the one hand and LR and $\lrgen$ on the other. Finally, we see how the
specific subjettiness ratios $\tau_2/\tau_1$ and $\tau_3/\tau_2$
increase for the HR/SR images, because the splitting of hard central
pixels into two hard and collinear, now resolved pixels increases the
IR-safe subjet count. The ratio $\tau_3/\tau_2$ turns out to be one of
the hardest of the HR-patterns to learn, with the effect that the SR
version leads to slightly smaller values.  This implies that the SR
network does not generate quite enough splittings. Such a feature
could of course be improved, but any optimization has to be balanced
with the ability of the network to also describe jets with more than
just collinear splittings, as we will see in the next case.

\subsection{Performance in Top-Quark Jets}
\label{sec:jets_top}

The physics of top-quark, light-quark, and QCD jets is very
different. While for QCD-jets collinear and, to some degree, soft
splittings describe the entire object, top-quark jets include the two
electroweak decay steps.  Comparing the top-quark jets shown in
Fig.~\ref{fig:top-top} with the QCD jets in Fig.~\ref{fig:qcd-qcd} we
see this difference already from the jet images --- the top-quark jets
are much wider and their energy is distributed among more pixels. From
a SR point of view, this simplifies the task, because the network can
work with more LR-structures. Technically, the adversarial loss
becomes more important, and we can indeed balance the performance on
top-quark jets vs QCD jets using $\lambda_\text{adv}$.

Looking at the ordered constituents, the additional mass drop
structure is learned by the networks extremely well. The leading four
constituents typically cover the three hard decay sub-jets, and they
are described even better than in the QCD case. Starting with the 4th
constituent, the relative position of the LR and HR peaks changes
towards a more QCD-like structure, so the network starts splitting one
hard LR-constituent into hard HR-constituents. This is consistent with
the top-quark jet consisting of three well-separated patterns, where
the QCD jets only show this pattern for one leading constituent. We
also see that up to the 15th constituent, the massive top-quark jet
shows comparably distinctive patterns and only few empty pixels.

For the high-level observables, we first see that the SR network
shifts the jet mass peak by about 10~GeV and does well on the girth
$w_\text{PF}$, aided by the fact that the jet resolution has hardly
any effect on the jet size. As for QCD-jets, $C_{0.2}$ is no challenge
for the up-sampling.  Unlike for QCD-jets, $\tau_3/\tau_2$ is as
stable as $\tau_2/\tau_1$, because it is completely governed by the
hard and geometrically well-separated hard decays.

\subsection{High-level observable benchmarking}
\label{sec:jets_benchmark}

\begin{figure}[t]
\centering
\includegraphics[width=0.30\textwidth]{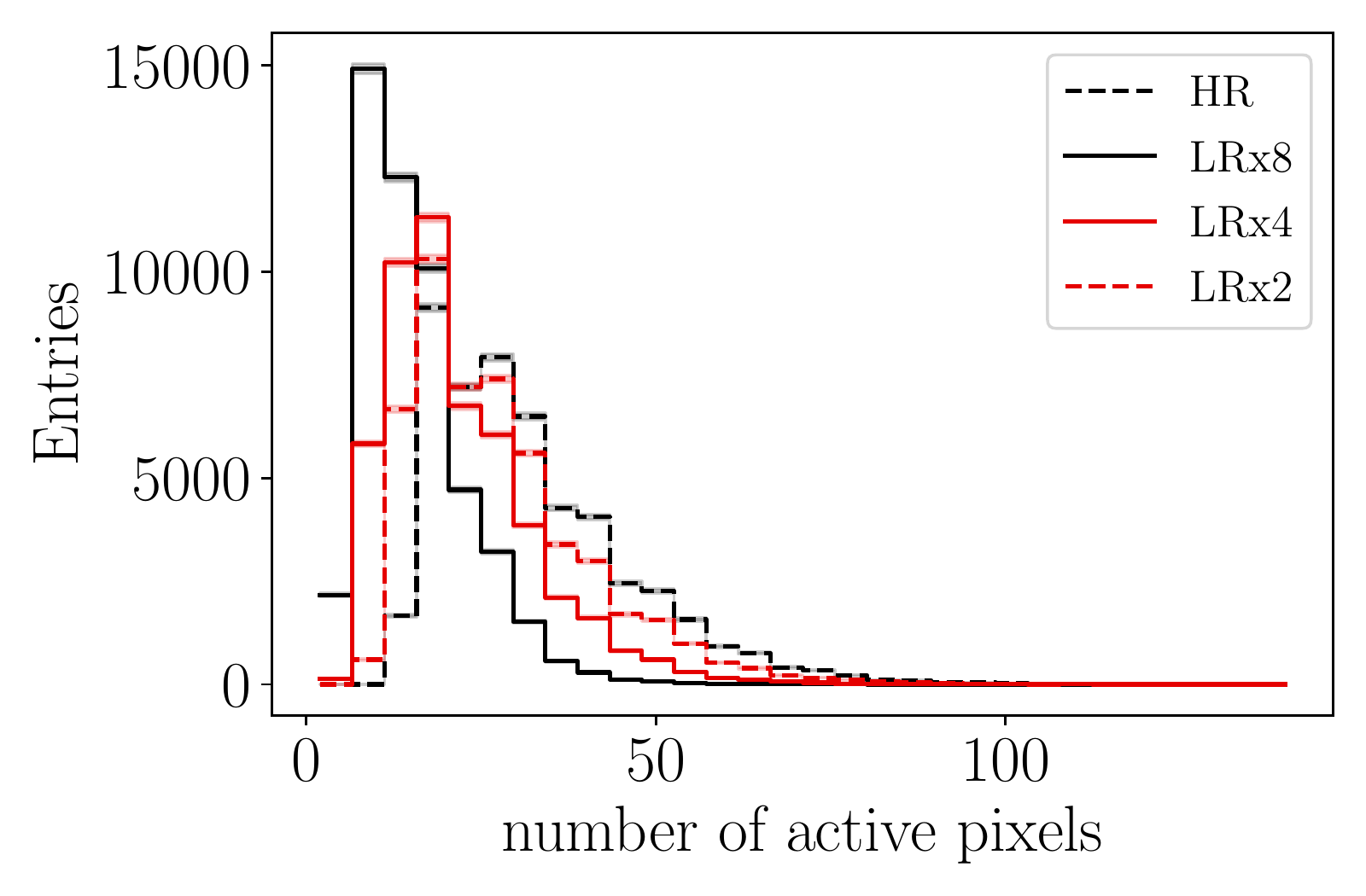}
\includegraphics[width=0.30\textwidth,page=1]{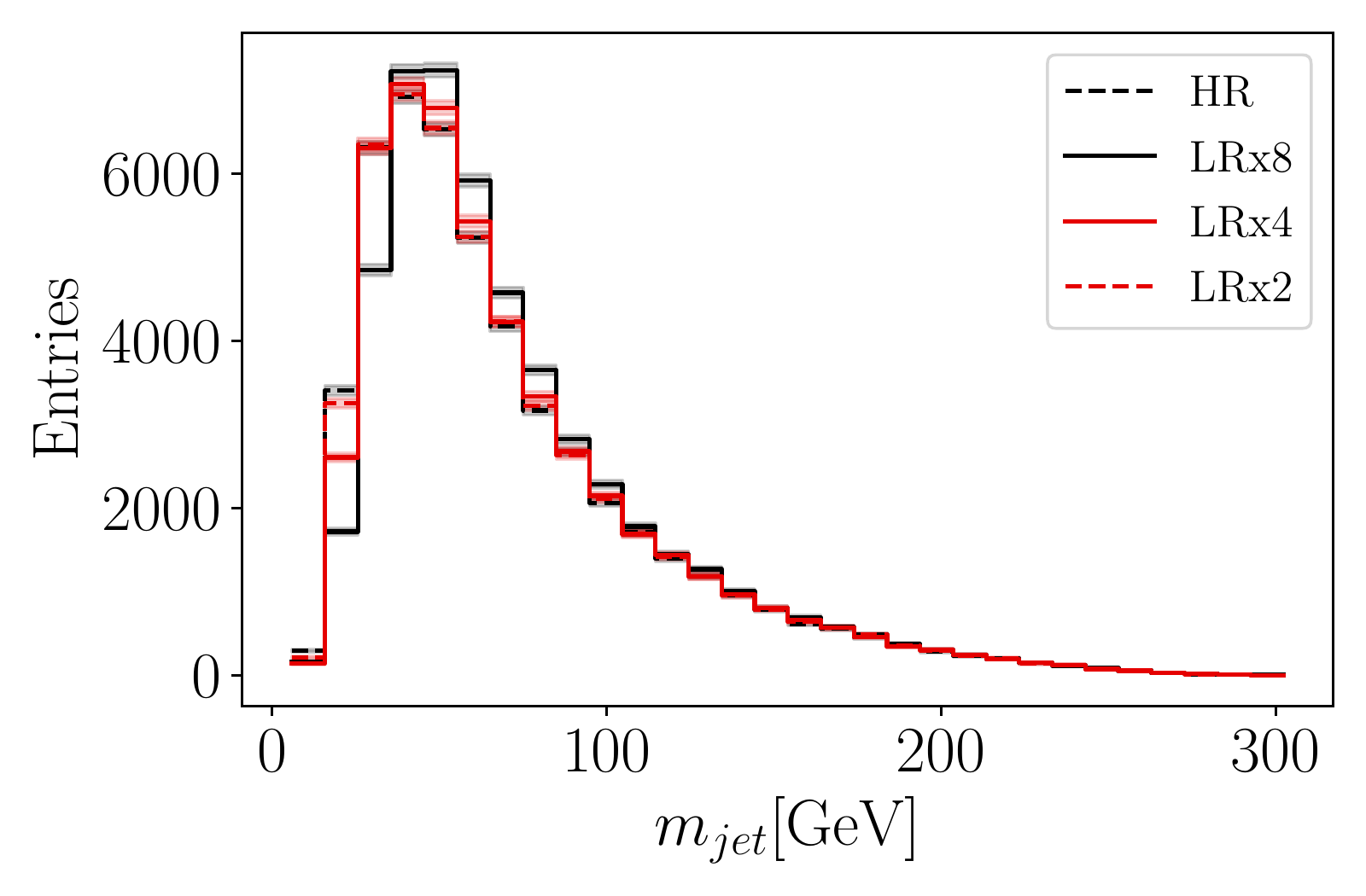}
\includegraphics[width=0.30\textwidth,page=4]{figures/qcdjetvarL8L4L2}\\
\includegraphics[width=0.30\textwidth,page=5]{figures/qcdjetvarL8L4L2}
\includegraphics[width=0.30\textwidth,page=2]{figures/qcdjetvarL8L4L2}
\includegraphics[width=0.30\textwidth,page=3]{figures/qcdjetvarL8L4L2}\\
\includegraphics[width=0.30\textwidth]{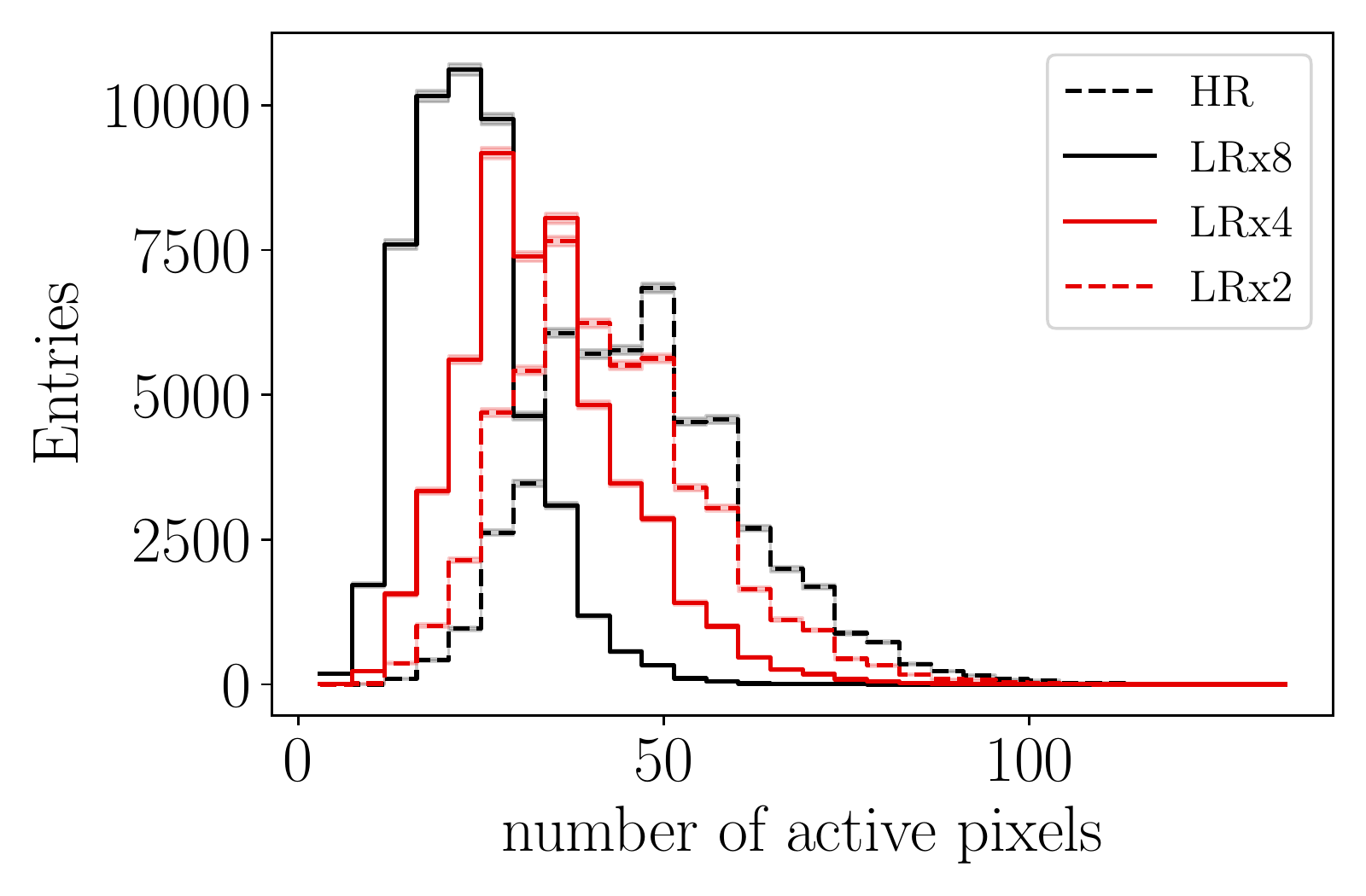}
\includegraphics[width=0.30\textwidth,page=1]{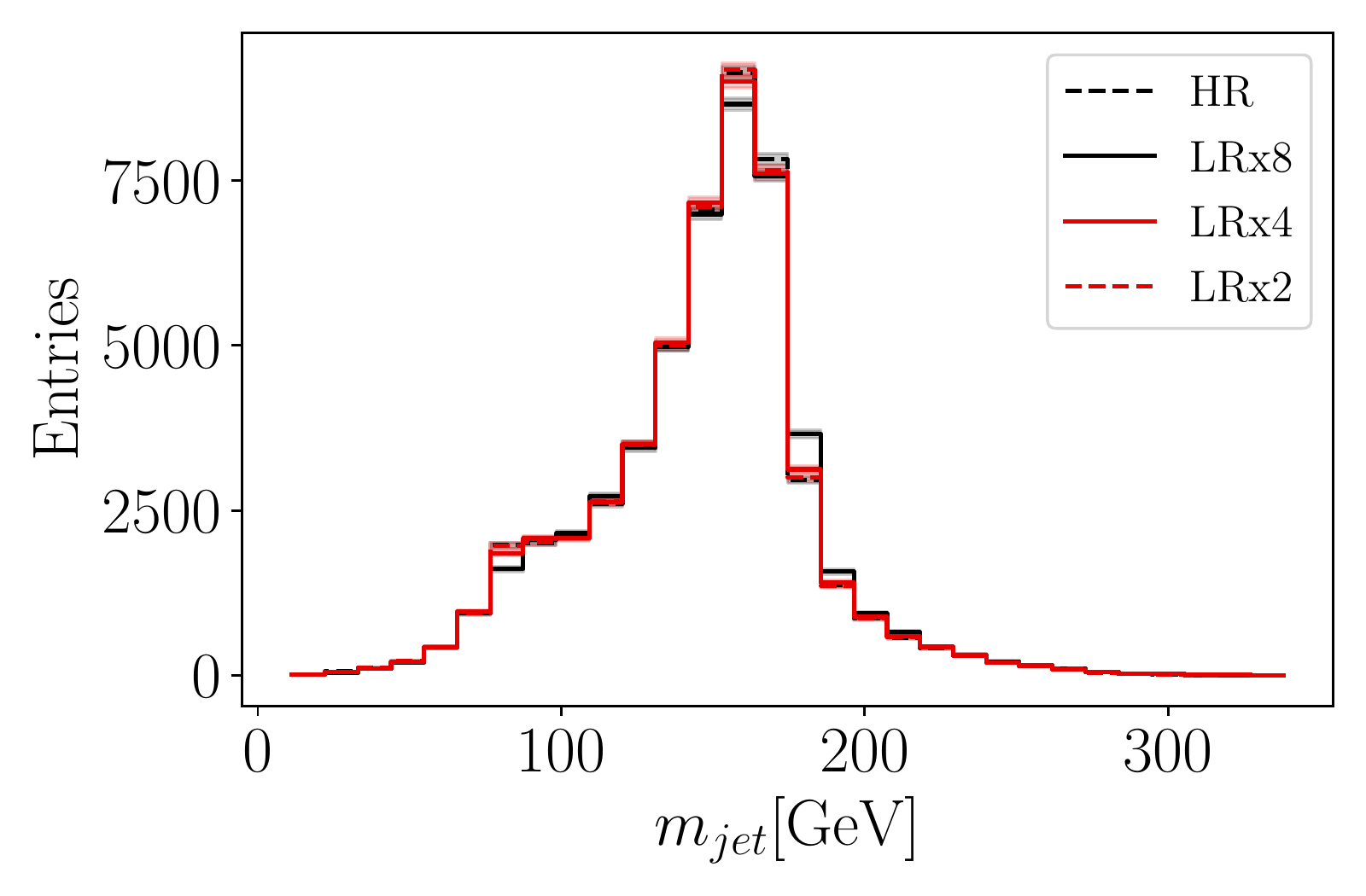}
\includegraphics[width=0.30\textwidth,page=4]{figures/topjetvarL8L4L2}\\
\includegraphics[width=0.30\textwidth,page=5]{figures/topjetvarL8L4L2}
\includegraphics[width=0.30\textwidth,page=2]{figures/topjetvarL8L4L2}
\includegraphics[width=0.30\textwidth,page=3]{figures/topjetvarL8L4L2}\\[-4mm]
\caption{Distribution of the number of active pixels and high-level observables 
  $m_\text{jet}$, $\tau_2/\tau_1$, $\tau_3/\tau_2$, $w_\text{pf}$, and 
  for images down-scaled by factors 2, 4, and 8. Results are shown for QCD 
  jets (top rows) and top-quark jets (bottom rows).}
\label{fig:hlo_steps}
\end{figure}

To understand the physics impact of our super-resolution approach and the possible 
gains, we inspect distributions of the benchmark high-level observables defined in Eq.\eqref{eq:jetobs} 
and the subjettiness ratios $\tau_{1,2,3}$. In a first step, we compare the one-step 
by a factor eight with a more continuous up-sampling.
In Fig.~\ref{fig:hlo_steps} we see that the three
different down-scaling steps indeed interpolate between the full HR
and LR jets smoothly. While the maximum in the number of active pixels
shifts almost linearly, the jet mass is stable and not affected much by down- and 
up-sampling at all. The $p_\textrm{T}$-weighted girth is only affected for 
the collimated QCD jets, similar to the subjettiness ratio $\tau_2/\tau_1$.  In
contrast, the correlator $C_{0.2}$ indicates a continuous loss in resolution for 
top-quark jets, as does the subjettiness ratio $\tau_3/\tau_2$.

\begin{figure}[t]
\centering
\includegraphics[width=0.45\textwidth,page=1]{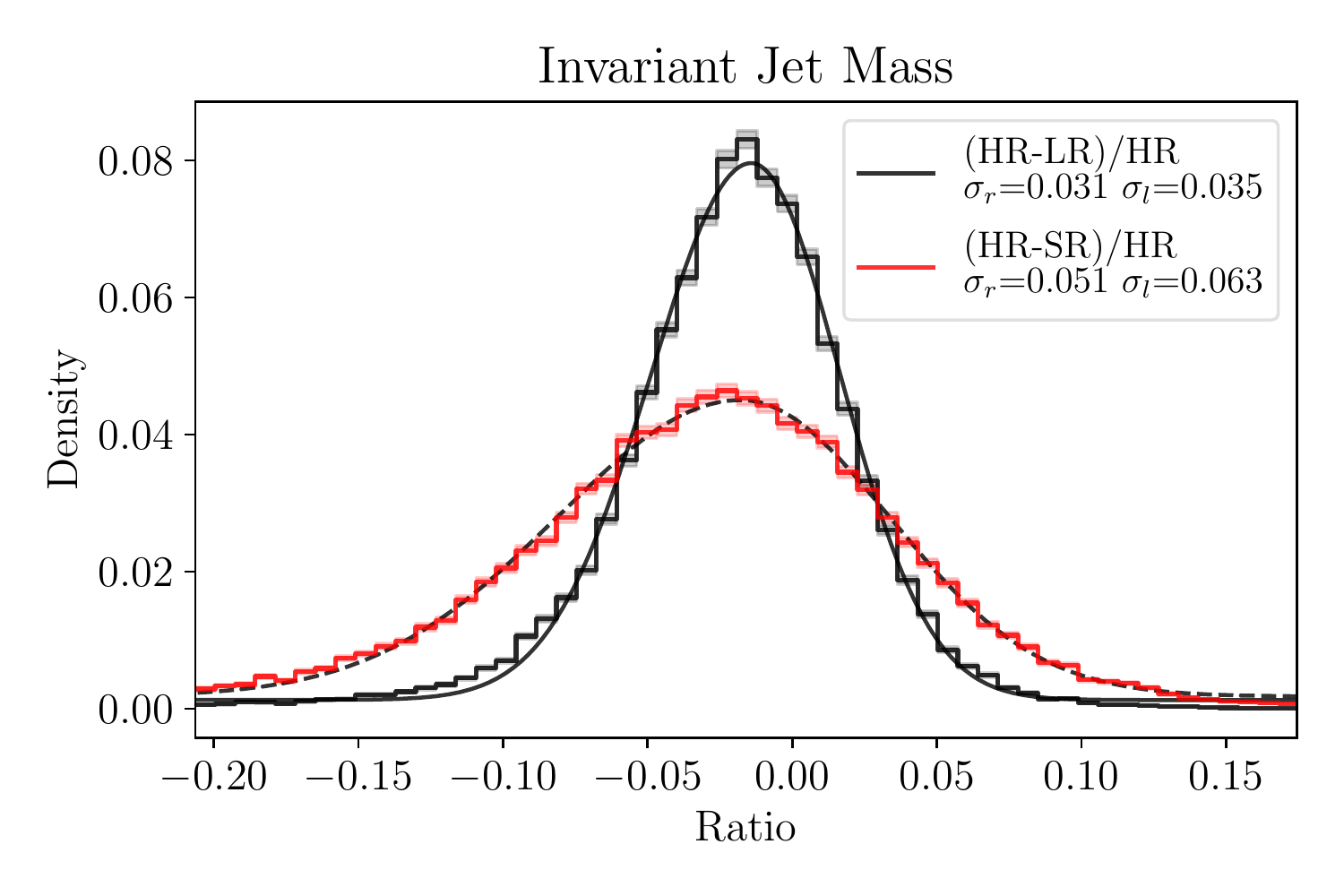}
\includegraphics[width=0.45\textwidth,page=2]{figures/relative_difference} \\
\includegraphics[width=0.45\textwidth,page=3]{figures/relative_difference}
\includegraphics[width=0.45\textwidth,page=4]{figures/relative_difference} \\
\includegraphics[width=0.45\textwidth,page=5]{figures/relative_difference}
\includegraphics[width=0.45\textwidth,page=6]{figures/relative_difference} \\
\includegraphics[width=0.45\textwidth,page=7]{figures/relative_difference}
\includegraphics[width=0.45\textwidth,page=8]{figures/relative_difference}
\\[-4mm]
\caption{Distribution of the relative difference in a given high-level observable for per-jet top-jet triplets LR, SR, and HR. In the upper panels we show the normalized deviations with a fitted double sided gaussian where the standard deviation is given by $\sigma(x)= \sigma_{l}$ for $x< \mu$ and $\sigma(x)= \sigma_{r}$ otherwise.
While in the lower panels we show ratios of deviations from the LR truth with a fitted sigmoid function. The quoted value $p$ gives the fraction of jet triples for which SR is closer to the  HR truth than LR truth is to HR truth, indicating that SR has added relevant information.}
\label{fig:perjet}
\end{figure}

The same high-level observables also allow us to tackle the key question,  whether  up-sampling adds information for an individual jet. The source of the additional information comes from trends in jet properties the network has learned for top-quark or QCD jets, rather than a magic recovery of lost resolution.  We remind ourselves the generative network is an inherently stochastic process, which provides a single example from a set of possible higher-resolution jets consistent with the observed lower-resolution jet, not a guess as to the true high resolution jet. Even a perfect network would display significant differences between generated jets and the single true jet. Instead, we can evaluate the information added by comparing how well the super-resolution jets estimate the true high-resolution jet information and compare it to how well the low-resolution jets estimate that same quantity.  Specifically, for a given observable, we examine the distribution of the quantities:
\begin{align}
\frac{\text{HR} - \text{LR}}{\text{HR}}
\qquad \text{and} \qquad
\frac{\text{HR} - \text{SR}}{\text{HR}} \; .
\end{align}
Figure~\ref{fig:perjet}  illustrates the relative per-jet performance for LR and SR jets. First, we confirm our observation from Fig.~\eqref{fig:hlo_steps} that the variation 
in the jet mass from down-sampling and from SR up-sampling is small (the distribution peaks near 0). The relative deviation
to the HR truth is between 5\% and 10\%, with an ever so slight bias towards larger jet masses, also 
visible already in Fig.~\eqref{fig:hlo_steps}. The increase in the width of the
differences for SR is just the effect of our stochastic generative approach for an 
essentially insensitive high-level observable. We note that the introduction of this extra source of stochasticity may pose challenges for analyses in which the precision of invariant jet mass is central. 

For the 2-subjettiness $\tau_2$ the variation is slightly larger than for the jet mass, and
both differences have similar widths, so the  information loss from down-sampling 
is almost compensated by the SR up-sampling. Moreover, down-sampling generates a visible 
bias, while the mean of the difference $\text{HR} - \text{SR}$ is nicely centered around
zero. Correlating the two deviations we indeed find that in 54\% of all jets the SR 
image is closer to the HR truth than the down-sampled image. This pattern is enhanced for 
the 3-jettiness $\tau_3$, where the effect of down-sampling increases in width and in bias.
In contrast, the super-resolution difference $\text{SR} - \text{HR}$ remains at negligible 
bias, and the increase in width is under better control. Consequently, in the lower 
panel we see that SR adds information for 61\% of all events already for our relative 
naive and un-tuned approach.

From Fig.~\eqref{fig:hlo_steps} we can hope for a dramatic impact of the super-resolution 
algorithm for $C_{0.2}$, because in spite of the $p_\text{T}$-weighting in the observable,
the information loss from down-sampling is big. Indeed, the bias in the difference 
$\text{HR} - \text{LR}$ becomes of order one, combined with a misleadingly small 
width of the distribution. The strong peak around an actually wrong value of $C_{0.2}$
implies that down-sampling has indeed lost a significant amount of information. The 
difference $\text{HR} - \text{SR}$ is broader than the difference from down-sampling, 
but as before the bias is better controlled. In the lower panel we confirm that for 
this non-trivial correlator the SR up-sampling adds significant information about
generic jet features and leads to an improvement for 74\% of all jets. 

The above analysis serves to illustrate that SR indeed adds generic QCD information which translates into established and relevant physics properties. However, while many of the SR variables show measured improvement over LR performance, the benefits of this increased information must be weighed by the increased stochasticity in other variables (i.e. Invariant Jet Mass). The utility of this method therefore depends heavily on the goals and scope of the particular analysis at hand. Future work should make this benchmarking systematic and investigate whether we can further improve the performance of our SR network on crucial correlators.
\subsection{Model dependence}
\label{sec:jets_model}

\begin{figure}[t]
\centering
\includegraphics[width=0.30\textwidth,page=2]{figures/TOPev_qcd_M2}
\includegraphics[width=0.30\textwidth,page=3 ]{figures/TOPev_qcd_M2}
\includegraphics[width=0.30\textwidth,page=4 ]{figures/TOPev_qcd_M2}\\
\includegraphics[width=0.30\textwidth,page=5]{figures/TOPev_qcd_M2}
\includegraphics[width=0.30\textwidth,page=6 ]{figures/TOPev_qcd_M2}
\includegraphics[width=0.30\textwidth,page=7 ]{figures/TOPev_qcd_M2}\\
\includegraphics[width=0.30\textwidth,page=8]{figures/TOPev_qcd_M2}
\includegraphics[width=0.30\textwidth,page=9 ]{figures/TOPev_qcd_M2}
\includegraphics[width=0.30\textwidth,page=10 ]{figures/TOPev_qcd_M2}\\
\includegraphics[width=0.30\textwidth,page=1]{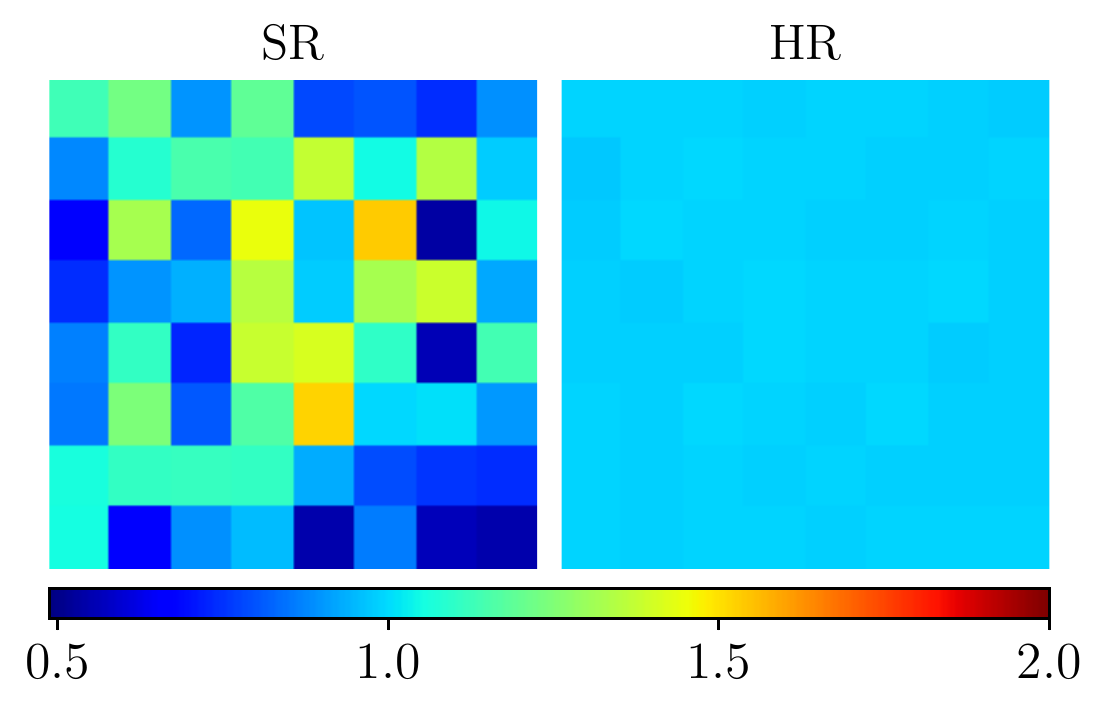}
\includegraphics[width=0.30\textwidth,page=11 ]{figures/TOPev_qcd_M2}
\includegraphics[width=0.30\textwidth,page=14 ]{figures/TOPev_qcd_M2}\\
\includegraphics[width=0.30\textwidth,page=15]{figures/TOPev_qcd_M2}
\includegraphics[width=0.30\textwidth,page=12 ]{figures/TOPev_qcd_M2}
\includegraphics[width=0.30\textwidth,page=13 ]{figures/TOPev_qcd_M2} \\[-4mm] 
\caption{Demonstration of the performance of a network trained on QCD
  jets and applied to top-quark jets.  Top left are averages of the HR
  and SR images, followed by distributions of the square-root of the
  energy of leading pixels, sub-leading, etc. Also shown are average
  ($f \times f$)-patches for the SR and the HR images, and
  distributions of high-level jet observables, see text for
  definitions. The zero-bin in energy collects jets with too few
  entries.}
\label{fig:qcd-top}
\end{figure}

\begin{figure}[t]
\centering
\includegraphics[width=0.30\textwidth,page=2]{figures/QCDev_top_M2}
\includegraphics[width=0.30\textwidth,page=3 ]{figures/QCDev_top_M2}
\includegraphics[width=0.30\textwidth,page=4 ]{figures/QCDev_top_M2}\\
\includegraphics[width=0.30\textwidth,page=5]{figures/QCDev_top_M2}
\includegraphics[width=0.30\textwidth,page=6 ]{figures/QCDev_top_M2}
\includegraphics[width=0.30\textwidth,page=7 ]{figures/QCDev_top_M2}\\
\includegraphics[width=0.30\textwidth,page=8]{figures/QCDev_top_M2}
\includegraphics[width=0.30\textwidth,page=9 ]{figures/QCDev_top_M2}
\includegraphics[width=0.30\textwidth,page=10 ]{figures/QCDev_top_M2}\\
\includegraphics[width=0.30\textwidth,page=1]{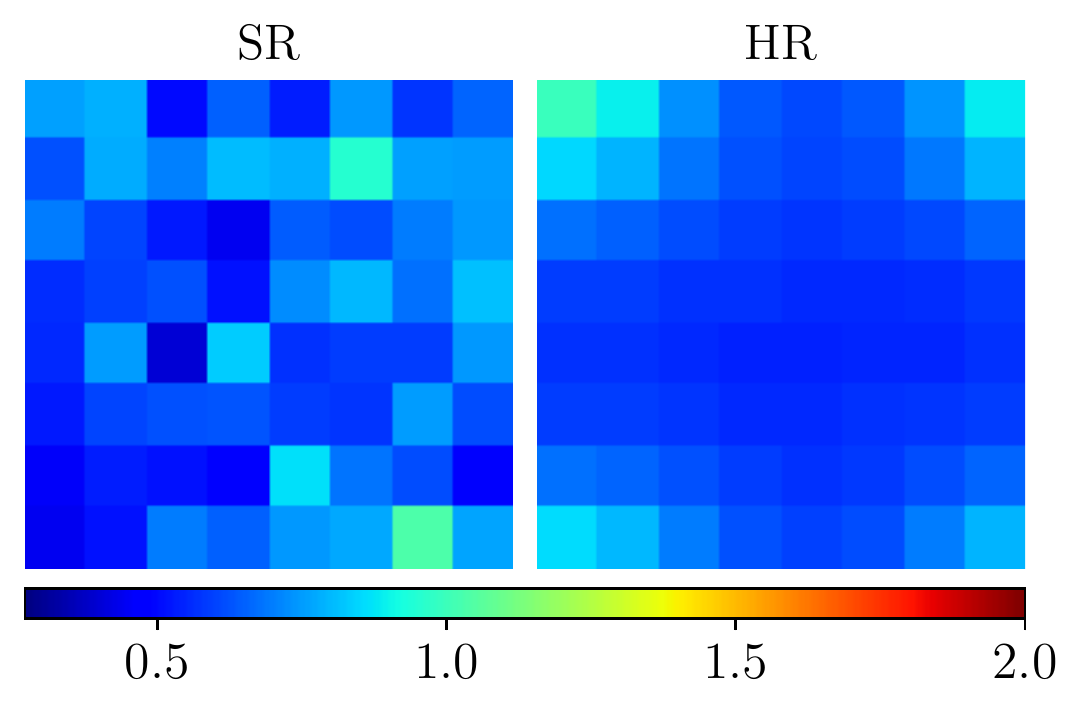}
\includegraphics[width=0.30\textwidth,page=11 ]{figures/QCDev_top_M2}
\includegraphics[width=0.30\textwidth,page=14 ]{figures/QCDev_top_M2}\\
\includegraphics[width=0.30\textwidth,page=15]{figures/QCDev_top_M2}
\includegraphics[width=0.30\textwidth,page=12 ]{figures/QCDev_top_M2}
\includegraphics[width=0.30\textwidth,page=13 ]{figures/QCDev_top_M2} \\[-4mm]
\caption{Demonstration of the performance of a network trained on
  top-quark jets and applied to QCD jets.  Top left are averages of
  the HR and SR images, followed by distributions of the square-root
  of the energy of leading pixels, sub-leading, etc. Also shown are
  average ($f \times f$)-patches for the SR and the HR images, and
  distributions of high-level jet observables, see text for
  definitions. The zero-bin in energy collects jets with too few
  entries.}
\label{fig:top-qcd}
\end{figure}

The ultimate goal for jet super-resolution is to learn jet structures
in general, such that SR images can be used to improve multi-jet
analyses. In practice, a network could then be trained on some kind of
representative jet sample. In our case, the QCD jets and top-quark
jets are extremely different, and we further amplify this effect by
training the models on one sample and applying them to the other. This
gives an example of a large model dependence and allows us to
understand the behavior by comparing with the correctly assigned data
sets.

In Fig.~\ref{fig:qcd-top}, we show the results from the network
trained on QCD jets, now applied to LR top-quark jets. Interestingly,
the network generates all the correct patterns for the ordered
top-quark jet constituents, albeit with a slightly reduced precision
for instance for the 15th constituent. Similarly, the patches still do
not include unwanted visible patterns, but are slightly more noisy.

Finally, in Fig.~\ref{fig:top-qcd} we show the results from the
network trained on top-quark jets, but applied to LR QCD-jets. In a
detailed comparison with Fig.~\ref{fig:qcd-qcd}, we see that the
network does not generate the more challenging QCD patterns out of the
narrow central pixel set. It starts to fail already for the first and
second constituents, and works slightly better for the 7th constituent
in the transition region before correctly reproducing the soft noise
patterns. In the distributions of high-level observables, the problem
is most evident in $\tau_2/\tau_1$. Here the training on the top-quark
sample pushes the SR QCD-image towards larger values or higher jet
multiplicities.  This reflects the broader structure of the training
sample with its generally larger values of $\tau_2/\tau_1$.

\begin{figure}[!ht]
\centering
\includegraphics[width=0.4\textwidth,page=2]{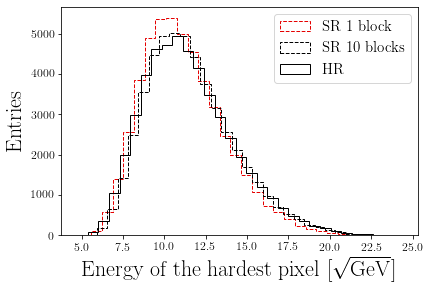}
\includegraphics[width=0.4\textwidth,page=3 ]{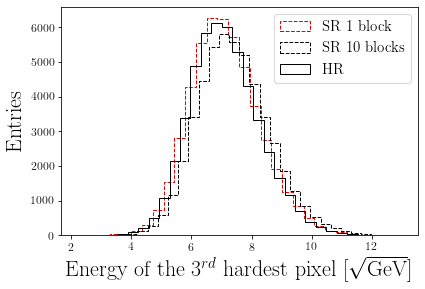} \\
\includegraphics[width=0.4\textwidth,page=4]{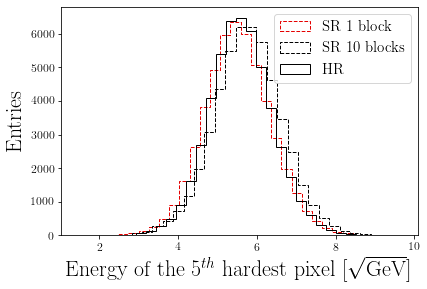}
\includegraphics[width=0.4\textwidth,page=5 ]{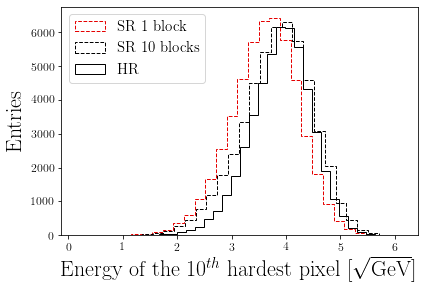} \\
\includegraphics[width=0.4\textwidth,page=6]{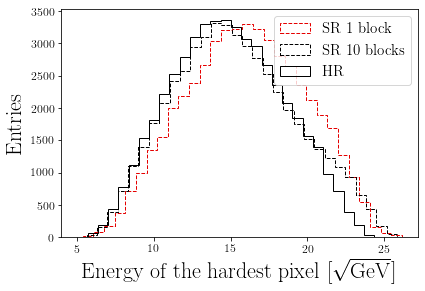}
\includegraphics[width=0.4\textwidth,page=7 ]{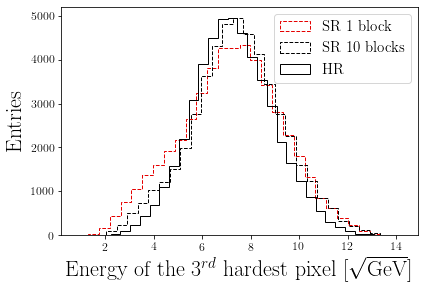} \\
\includegraphics[width=0.4\textwidth,page=8]{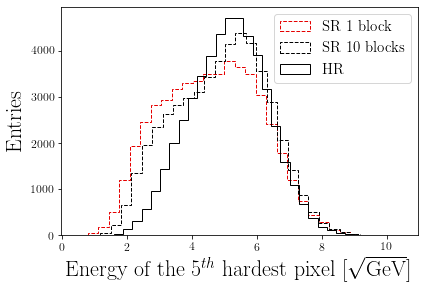}
\includegraphics[width=0.4\textwidth,page=9 ]{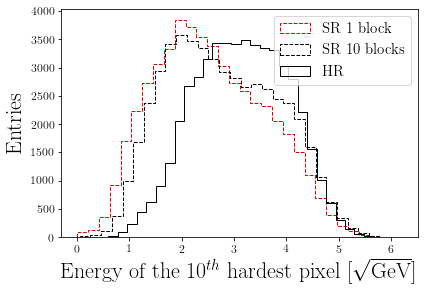}\\
\caption{Demonstration of the performance of a reduced complexity (1
  RRDB block) network compared to a more complex network (10 RRDB
  blocks), for networks trained on and applied top-quark jets (upper)
  and QCD jets (lower). Shown are distributions of the square-root of
  the pixel energies for the true high resolution image (HR) and super
  resolution images generated by the reduced and standard complexity
  network.}
    \label{fig:slim_histograms}
\end{figure}

\subsection{Network Complexity Reduction}
\label{sec:jets_slim}

The flexibility of deep networks often comes at a cost of
complexity. This complexity, in the form of a large number of layers
and nodes, means a large number of parameters must be optimized during
training. This hyper-flexibility can lead to undesirable side-effects
that ultimately hurt its utility especially when it comes to
systematic studies. A network with fewer parameters, which achieves
the same performance, will be more efficient to train, faster to
evaluate, less prone to over-fitting and more likely to
generalize. For these reasons, we aim to determine the minimal
necessary complexity of our GANs by systematically reducing the number
of layers until performance is impacted.

Most of our network complexity resides in the core of the
super-resolution GANs, which comprises the residual-in-residual dense
blocks (RRDBs), each of which includes 15 convolutional layers. In
this section we experiment with a smaller number of blocks, but the
same network architecture. In Fig.~\ref{fig:slim_histograms} we
compare pixel energy distributions for SR images generated by the
reduced-complexity network to those generated by the network described
earlier. In the first panels we see that for top-quark jet even a
single-block network is able to extract the truth features very
well. The remaining challenge is to properly describe the softer
pixels, just as we see for the full network in
Fig.~\ref{fig:top-top}. In the second set of panels in
Fig.~\ref{fig:slim_histograms} we show the corresponding result for a
network trained on and applied to QCD jets. As expected, the network
task is much more challenging because of the smaller number of
available LR-pixels and the much more focussed structure of QCD
jets. Similar to the full network results shown in
Fig.~\ref{fig:qcd-qcd}, the slim network does not push the energy for
the softer pixels to the full truth values, but gets stuck at a
slightly softer spectrum.

\begin{figure}[!ht]
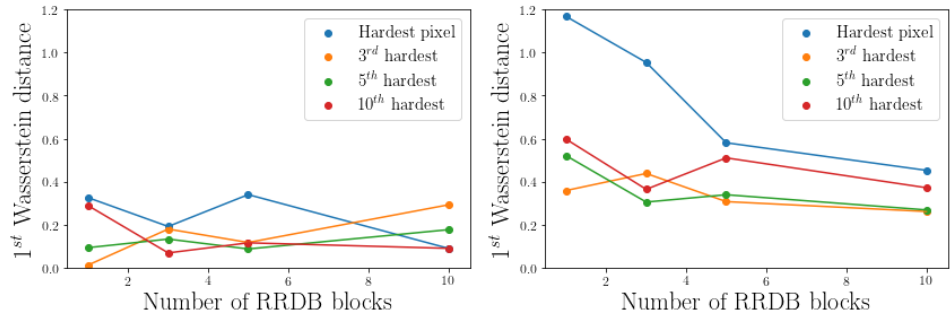

    \centering
    \includegraphics[width=0.4\textwidth,page=2]{figures/top_blocks_vs_10_hardest_HR}
    \includegraphics[width=0.4\textwidth,page=3 ]{figures/qcd_blocks_vs_10_hardest_HR}
    \caption{ Dependence of the performance of super resolution
      networks on the number of internal RRDB blocks (See
      Fig.~\ref{fig:network}). Performance is measured via the one
      dimensional Wasserstein distance between the distribution of
      quantities over true high-resolution images and the
      super-resolution images. Quantities examined are the energy of
      the leading pixel, subleading, etc.  Left (right) shows results
      for networks trained on top-quark (QCD) jets and applied to
      top-quark (QCD) jets.}
    \label{fig:blocksVsNHardestTop}
\end{figure}

To illustrate the super-resolution network performance we compute the
first Wasserstein distance between the true HR images and the SR
images. In Fig.~\ref{fig:blocksVsNHardestTop} we show this Wasserstein
distance as a function of the number of RRDBs for top-quark jets
(left) and QCD jets (right). The global scale of Wasserstein distance
values reflects the fact that top-quark jets are better described by
all networks, regardless of the number of RRDBs. As a matter of fact,
here the performance improvement from more RRDBs is almost completely
covered by the fluctuations from different network initializations and
runs. In contrast, the more challenging QCD jets show a significant
improvement with an increased network complexity. Interestingly, for
both top-quark and QCD jets, the performance improvement is not
visibly related to, for instance, hard vs soft pixels.
We also emphasize that the larger network complexity required by QCD
jets is in contrast to the complexity of the actual jets. While the
top-quark jets combine massive decay and QCD splitting patterns, the
physics principles behind the QCD jets are much simpler, so the
required complexity of the super-resolution network is not driven by
the complexity of the underlying objects, but by the effect of the
reduced resolution.

\section{Outlook}

Jet physics in terms of low-level observables and with the help of
deep networks defines many new opportunities in jet physics and jet
measurements at the LHC. For jet classification, or jet tagging,  deep networks typically outperform established high-level
approaches.

In this paper, we propose a new application of deep learning to jet
physics: jet super-resolution, which aims to overcome the limitations of detector resolution and allow for deeper analysis of jet data from ATLAS and CMS.
 Super-resolution
networks can provide additional information, and hence improved resolution, by encoding our knowledge about jet physics in a
generative network.

Our results demonstrate that a super-resolution network can indeed reproduce
high-resolution jet images of top-quark jets and QCD jets when trained on these samples. We illustrated the performance
of the super-resolution networks using images, low-level observables,
and high-level observables. The more challenging test of the generality of the network is evaluated by applying a network trained on one sample to jets from the other sample. We confirmed that our
super-resolution network exhibits the necessary model independence to
be applied to different kinds of jets. 
This will allow us to train jet super-resolution networks on mixed samples 
and avoid complications for instance with the poorly defined 
separation of quark and gluon jets in a QCD sample.

While the main focus of our study was to show that the technique of 
image super-resolution works reliably on LHC jets, we already showed that it can be used to enhance jet measurements in regions with poor calorimeter performance.
Additionally, we showed that the necessary complexity of the network depends on the source of the jets. Interestingly, equivalent performance on top-quark jets can be achieved with far fewer parameters than QCD jets, despite the former having greater complexities in the underlying physics mechanisms. Such knowledge is helpful in efficiently allocating computational resources when analyzing experimental jet data.

\begin{center} \textbf{Acknowledgments} \end{center}

We would like to thank Monica Dunford and Hans-Christian
Schultz-Coulon for the experimental encouragement.  The research of AB
is supported by the Deutsche Forschungsgemeinschaft (DFG, German
Research Foundation) under grant 396021762 -- TRR~257 \textsl{Particle
  Physics Phenomenology after the Higgs Discovery}.  DW is supported
by the Department of Energy, Office of Science. JNH acknowledges
support by the National Science Foundation under grants DGE-1633631
and DGE-1839285. Any opinions, findings, and conclusions or
recommendations expressed in this material are those of the author(s)
and do not necessarily reflect the views of the National Science
Foundation.

\bibliography{literature}
\end{document}